\documentclass[useAMS,usenatbib,usegraphicx]{mn2e}

\title{Detecting compact dark matter in galaxy clusters via
gravitational microlensing: A2218 \& A370}

\author[Tuntsov et al.]{
A.V. Tuntsov$^{1}$\thanks{E-mail:tyomich@physics.usyd.edu.au}, 
G.F. Lewis$^{1}$\thanks{E-mail: gfl@physics.usyd.edu.au},
R.A. Ibata$^{2}$\thanks{E-mail: ibata@pleiades.u-strasbg.fr} \&
J.-P. Kneib$^{3}$\thanks{E-mail: kneib@ast.obs-mip.fr},\\
$^{1}$A29, School of Physics, University of Sydney, NSW 2006, Australia\\
$^{2}$Observatoire de Strasbourg, 11, rue de l'Universite, F-67000, Strasbourg,
France\\
$^{3}$Observatoire Midi-Pyr\'{e}n\'{e}es, 14 Av. E. Belin, 31400 Toulouse, France\\
on leave at Caltech, Astronomy, 105-24, Pasadena, CA 91125, USA
}

\begin{document}

\date{Accepted 2004 June 10. Received 2004 June 8; in original form 2004 March 31}

\pagerange{\pageref{firstpage}--\pageref{lastpage}} \pubyear{2004}

\maketitle

\label{firstpage}
\begin{abstract}
After  decades of  searching, the  true  nature of  dark matter  still
eludes us.  One  potential probe of the form of  dark matter in galaxy
clusters is to  search for microlensing variability in  the giant arcs
and  arclets. In this paper, a simple method is introduced to characterize pixel variability in the limit of high optical depth to microlensing. Expanding on  earlier  work, the
expected microlensing signal for  two massive clusters, A2218 \& A370 is calculated.
It is found that the microlensing signal depends sensitively upon the
mix of smooth and compact dark matter in the cluster. Comparison of two deep exposures taken with {\it James Webb Space Telescope} or two hour long exposures taken with a 30-metre class telescope in two epochs separated by a few years will possibly detect about a few dozen pixels which show strong variability due to microlensing at $5\sigma$ level, revealing wealth of information on the microlensing population.
\end{abstract}


\section{Introduction}

There is significant  evidence that most of the  energy density in the
Universe resides in the forms  yet unknown, but the physical nature of
this dark  matter is  an issue of  live debate  and is far  from being
understood  \citep{lahavliddle, munnos}. Typical  explainations invoke
either  compact  objects,  such  as primordial  black  holes,  stellar
remnants  or   planets,  or   continuous  material  such   as  weakly
interacting  massive  particles,   unseen  hydrogen  or  more  exotic
explainations.  Potential clues  to the nature
of  dark matter  have recently  been uncovered  with the  detection of
neutrino mass  and observations of compact massive  objects thought to
be mostly white dwarfs in Galactic neighbourhood (e.g., \citealt{fukuda,
alcock}).   However, given  their mass  neutrinos cannot  contain more
than $\sim$13 per cent  of the  dark matter mass budget  \citep{hu}.  Furthermore,
the halo  microlensing programs constrain compact objects  in the mass
range of $(10^{-7}  - 10^{-1}) M_{\odot}$ to making  no more than 10 per cent
of the Galactic halo \citep{sadoulet}.

Gravitational lensing  has allowed the detailed  reconstruction of the
projected mass  distribution in galaxy  clusters (e.g., \citealt{ebbels}).
Unfortunately such analysis does not probe the  fundamental nature of
dark  matter.   However, although  deflection  angles  due to  compact
objects in  the dark  matter are negligible  their derivatives  may be
substantial which  means that although the granularity  of dark matter
cannot   change  the   morphology   of  lensed   systems,  this   {\it
microlensing}   can   have   significant   impact  on   the   observed
fluxes \citep{liebes,  paczynski}.   As  the  mutual  configuration  of
sources and  lenses change with  time, this results in  variability of
the observed flux.

Several microlensing  programs have been proposed  and undertaken, the
most  successful of  them being  the microlensing  experiments towards
Galactic  bulge  and  Magellanic   Clouds  by  MACHO,  EROS  and  OGLE
collaborations \citep{alcock, lasserre, udalski}.  Another approach is
pixel lensing when the search focuses not on individual sources but on
a large number of stars seen as a single pixel of the image. Events of
strong lensing  of a single  star can be  detected in the  pixel light
though they  never dominate  it.  Such events  have been  discussed in
connection  with  the galaxies  M31  and  M87 \citep{crotts,  baillon,
gould}, and a few observational programs monitoring these galaxies 
have been implemented \citep{crottstomaney, ansari, riffeser, ebaltz}.

Further  away from  the Galaxy,  it has  been proposed  to  search for
microlensing-induced  variability  of   quasars  seen  through  galaxy
clusters \citep{walkerireland,  tadros}. This has  been undertaken for
the  Virgo cluster,  although  its close  proximity  ensures that  the
optical depth to microlensing is low.  Although analysis of the quasar
variability based on a  long observation series by \citet{hawkins} let
him conclude  that the  dark matter may  be dominated  by Jupiter-mass
microlenses,  this   idea  is  not  widely  supported.    One  of  the
difficulties here  is that quasars  are expected to  possess intrinsic
variability   \citep{zackrisson}   which   significantly   complicates
microlensing studies. Totani (2003)  proposed to explore the advantage
of the recent discovery of a galaxy cluster found just behind the rich
cluster  Abell  2152   but  this  has yet to  result  in  an
observational program.

Another  possibility  is  to   look  for  the  variations  in  surface
brightness  of  strongly  lensed   distant  galaxies  --  e.g.,  giant
gravitationally  lensed  arcs (Lewis \& Ibata 2001; Lewis, Ibata \& Wyithe 2000).  They  are
found in the  regions of large magnification and therefore high optical depth. Assuming that microlenses make up an appreciable fraction of the lensing mass, it means that in any instant all  the stars in a  pixel are subject  to strong microlensing
and  therefore an investigation  of the surface brightness  variability  is
effectively reduced to an investigation of  the behaviour of the sum of a
large number  of individual fluxes  with fairly well  known individual
statistical properties.

This study  extends the analysis  of \citet{gflibataw} and  provides a
general method for the analytic  calculation of variability patterns given
the convergence and shear values.  We apply it to the two well-studied
galaxy clusters  - Abell 370 and  Abell 2218.  In the  next section we
describe and justify  the method we use, and  then calculate the value
of  individual   microlensing-induced  variability  dispersion   as  a
function  of  standard microlensing  parameters.  Section 3  discusses
observational  prospects for  detecting this  sort of  variability. We
discuss our results in section 4.

\section[]{General method}

\subsection[]{Statistical approach}

The goal of this section is to learn to characterize the variability in pixels
of images of distant galaxies with a few parameters. We show that the high
optical depth and large numbers of stars forming the pixel allow one to reduce
this to a single parameter -- dispersion -- as the observed flux distribution 
can be approximated by the Gaussian function with good accuracy.

The major assumption we make  in our investigation is  that the fluxes of individual stars,
which vary as  stars and lenses move with respect to  each other, do so
independently. This is a natural assumption since individual stars in the source plane are distributed in a random manner. It might be worth considering the presence
of some form of agglomerations in pixels but this seems to be an unnecessary
complication at the present stage.

We consider statistical properties `as  static' -- i.e. on time scales
greater than typical individual variability scale given by the time of
Einstein-Chwolson  radius   crossing  ($r_e/v$),  and   therefore  are
interested  in the  probability  distribution function  of the  summed
pixel flux.  The typical fluxes in pixels forming images of strongly  lensed galaxies correspond to the range in luminosity of thousands to billions of solar luminosities  -  and therefore  contain  very  large numbers  of stars. This  fact along  with the  variability independence
immediately  suggests using  the central  limit theorem  to  infer
the statistical properties of the pixel flux.

Indeed,  let  the  intrinsic  (unlensed)  fluxes  of  the  pixel
population stars be $\{L_i\}, i=1..N$, N being the number of stars in the pixel.  If we neglect the fraction of
stars intrinsically variable  at the same level and  time scale as the
variability  caused by  the  microlensing,  the role  of  $L_i$ is  to
normalize the variability in magnification factor $\mu_i$ arising as
the source stars move through the magnification map:
\begin{equation}
	L_i^{obs}=L_i\mu_i \label{Liobs}
\end{equation}
and $\mu_i$ can be considered a random variable.

The probability distribution of  $\mu_i$ does depend on the individual
characteristics   of  the   source  --   mostly,  the   size   of  its
disc.  However,  as  we  will  see  in  the  following  section,  this
dependence  is  not strong  and,  to  show  the validity  of  Gaussian
approximations  it is enough  to assume  that $\mu_i$  distribution is
identical for all  $i$ and only depends on  (macro) lensing parameters
in the pixel.\footnote{We also assume that these parameters are constant
throughout the pixel and this is a natural assumptions in most cases.}

The flux observed in the pixel is given by
\begin{equation}
	L^{obs}|_{\{L_i\}}=\sum\limits_{i=1}^N L_i\mu_i \label{LobspriLi}
\end{equation}
and its average value is
\begin{equation}
	\overline{L^{obs}}|_{\{L_i\}}=\overline{\mu}\sum\limits_{i=1}^N L_i, \label{avLobs}
\end{equation}
where the bar  denotes the averaging  over the lensing configuration  at a given
convergence and shear  (assuming a sort of ergodic  hypothesis this is
equivalent to averaging over time).

Let us define $\delta L_i\equiv(\mu_i-\overline{\mu})L_i$ and consider
the deviation of $L^{obs}|_{\{L_i\}}$ from its average value:
\begin{equation}
\delta L|_{\{L_i\}}=\sum\limits_{i=1}^NL_i(\mu_i-\overline{\mu}) \label{deltaLpriLi}
\end{equation}
Let us  also define  the second and  third moments of the distribution in
$\mu$:   $\sigma^2_{\mu}\equiv\overline{(\mu-\overline{\mu})^2}$   and
$\beta\equiv\overline{|\mu-\overline{\mu}|^3}$.      Clearly,      the
corresponding moments for $\delta L_i$ equal $L_i^2\sigma^2_{\mu}$ and
$L_i^3\beta$.

According  to Lyapunov  theorem, the  actual  probability distribution
function   F  of   $\delta   L|_{\{L_i\}}$  tends   to  the   Gaussian
approximation
\begin{equation}
\Phi(\delta L)=\frac{1}{\sqrt{2\pi\sigma^2}}e^{-\frac{\delta L^2}{2\sigma^2}} \label{Gaussi}
\end{equation}
where we dropped $\{L_i\}$ subscript for clarity, with
\begin{equation}
\sigma^2\equiv\sigma^2_{\mu}\sum\limits_i L_i^2 \label{sigmaGaussi}
\end{equation}
The  accuracy  of  this  approximation  in  Kolmogorov  ($L_{\infty}$)
measure  $\rho(F,\Phi)\equiv\sup\limits_{x\in{\cal R}}|F(x)-\Phi(x)|$  is  not worse  than
$cA$, where
\begin{equation}
A\equiv\frac{\beta}{\sigma_{\mu}^3}\frac{\sum L_i^3}{(\sum L_i^2)^{3/2}} \label{accuBEC}
\end{equation}
\citep{berry,                        esseen}                       and
$c\equiv\frac{\sqrt{10}+3}{6\sqrt{2\pi}}\approx                   0.41$
\citep{chistjakov}.

Although the  value of  the second fraction  in $A$ may,  depending on
$\{L_i\}$, be as large as unity its typical value is of order $\langle
L_i^3\rangle/(\langle   L_i^2\rangle^{3/2}\sqrt{N})$,   where   the   angled
brackets denote averaging over the luminosity function (which, up to a distance- and band-dependent constant, is the distribution of intrinsic fluxes).

With        the       luminosity        function        given       by
\citet{jahreisswielen}\footnote{The work of \citet{jahreisswielen} presents V-band luminosities
which  are   of  rather   limited  interest  for   exact  cosmological
predictions where K-corrections are to be taken into account; however,
these  numbers provide  sensible  estimates for  the quantities  under
consideration.},  $\langle  L_i^3\rangle\approx  10^3 L_{\odot}^3$  and
$\langle L_i^2\rangle\approx 10 L_{\odot}^2$, therefore
\begin{equation}
cA\approx 10\frac{\beta}{\sigma_{\mu}^3}\frac{1}{\sqrt{N}} \,. \label{cAest}
\end{equation}
To estimate the values of  $\beta$ and $\sigma_{\mu}$ it is sufficient
to use a rather coarse `model' probability distribution density $p(\mu)$
which  is   normalized  to  unity  and  has   three  basic  properties
established theoretically:
\begin{enumerate}
\item $p(\mu)=0$ at $\mu\le 1$
\item      $\int\rmn{d}\mu\, p(\mu)      \mu      =      \mu_{th}$,
$\mu_{th}=|(1-\kappa^2)-\gamma^2|^{-1}$
\item $p(\mu)\sim \mu^{-3}$ at $\mu \gg 1$.
\end{enumerate}
The job is done by the following `model' $p(\mu)$:
\begin{equation}
	p(\mu)=\frac{2(\mu_o-1)^2(\mu-1)}{[(\mu-1)^2+(\mu_o-1)^2]^2} \,. \label{modelpmu}
\end{equation}
The second condition implies $\mu_o-1=2(\mu_{th}-1)/\pi$.

This distribution does not possess the second moment, let  alone the third, as a result of the (iii) property. However, the finite size
of the source places a cut-off $\mu_{max}$ at the high values of $\mu$
(for a single point mass $\mu_{max}$ is nearly inversely proportional to
the   source   size   as   was   shown   by   \citet{liebes}).   Since
$\mu_{max}\gg\mu_o$ it does not affect either norm or the first moment
and therefore with $\sigma_{\mu}^2\approx 2\mu_o^2\ln{(\mu_{max}/\mu_o)}\approx\mu_{th}^2\ln{(\pi\mu_{max}/2\mu_{th})}$
and $\beta\approx \mu_{th}^2\mu_{max}$ we have
\begin{equation}
	cA\approx\frac{10}{\sqrt{N}}\frac{\mu_{max}}{\mu_{th}\ln{(\pi\mu_{max}/2\mu_{th})}} \,. \label{cAhere}
\end{equation}
With   typical  for gravitationally lensed arcs values  of  $\mu_{th}\sim   10$,
$\mu_{max}\sim 100$  \citep{gflibataw} and $N  \sim 10^4 -  10^6$~in a
pixel $cA\la 0.01 - 0.1$. This means that when talking about a deviation
of  at   least  one  standard  value  $\sigma$,   for  which the Gaussian
probability  is  $\approx 0.15$,  one  can  be  sure that  the  actual
probability of such a deviation is not less than (5 -- 25) per cent.

Strictly speaking, the minimum magnification value for microlensing at high optical depth is greater than one used in (i) as was shown by \citet{schneider84}, and one could rather use some model value for this quantity \citep{bartschneider}.  However, this does not have much impact on the estimate of the validity of our approximation. Perhaps more important is that due to the value of minimum magnification which is greater than unity, Gaussian approximation clearly cannot hold exactly as it assigns non-zero probability to flux values below the minimum. However, this inconsistency is well inside the uncertainty of our method, given by~(\ref{cAhere}) and does not affect our results.

The initial task is thus  reduced to calculating the only parameter of
a centered  Gaussian distribution -- its dispersion. As $\{L_i\}$  is not
known  {\it  a priori}  (and  neither  it can  be  known  well {\it  a
posteriori}),  $\delta  L|_{\{L_i\}}$  is  to  be  averaged  over  all
possible  $\{L_i\}$.  This  can easily be done  by  considering  the
following three random variables:
\begin{equation}
\Delta L=\Delta L^{o}+\delta L \label{dvedelty}
\end{equation}
where    $\Delta   L=\sum\limits_i(\mu    L_i   -\overline{\mu}\langle
L_i\rangle)$,   $\Delta  L^{o}=\overline{\mu}\sum\limits_i(L_i-\langle
L_i\rangle)$ and  $\delta L$ is  the value of  interest. Here,
again the bar denotes  averaging in the $\mu$ domain while  angled brackets mean
averaging over $\{L_i\}$.

As these three quantities  are (nearly) Gaussian and uncorrelated (the
correlation  vanishes  when   averaging  over  $\mu$),  the  following
relation holds:
\begin{equation}
	\sigma^2_{\delta L}=\sigma^2_{\Delta L} - \sigma^2_{\Delta L^o} \,. \label{sigmadeltaL}
\end{equation}
Clearly, 
\begin{equation}
\sigma^2_{\Delta L^o} = \Bigr\langle\left(\overline{\mu}\sum\left(L-\langle L_i\rangle\right)\right)^2\Bigl\rangle = N\overline{\mu}^2\sigma^2_L \label{sigmadeltalo}
\end{equation}
where     $\sigma^2_L$    is  the   dispersion of individual flux \citep{jahreisswielen}.

For the second quantity we may write:
\begin{eqnarray}
\sigma^2_{\Delta L}&= &\overline{\left\langle\left(\sum\left(\mu L - \overline{\mu}\langle L_i\rangle\right)\right)^2\right\rangle} \label{sigmadeltalb}\\
 & = &N\overline{\left\langle\mu^2L_i^2-2\overline{\mu}\mu L_i\langle L_i\rangle+\overline{\mu}^2\langle L_i\rangle^2\right\rangle} \nonumber \\
 & = &N\left(\overline{\mu^2}\langle L_i^2\rangle-\overline{\mu}^2\langle L_i\rangle ^2\right)\nonumber \\
 & = &N\left(\sigma^2_{\mu}\langle L_i^2\rangle +\overline{\mu}^2\sigma^2_L\right) \nonumber
\end{eqnarray}
and therefore
\begin{equation}
\sigma^2_{\delta L}=N\langle L_i^2\rangle\sigma^2_{\mu} \,. \label{sigmaofinterest}
\end{equation}
The dispersion   of  the   quantity   $\delta  L/(N\overline{{\mu}}\langle
L_i\rangle)$ -- the relative fluctuation -- is thus given by:
\begin{equation}
\label{epsilondeltal}
\epsilon_{\delta L}^2=\frac{1}{N}\frac{\langle L_i^2\rangle}{\langle L_i\rangle^2}\frac{\sigma^2_{\mu}}{\overline{\mu}^2} \,.
\end{equation}
As $N$ is not known either we will just divide the observed flux
of  the pixel  $L_{obs}$ by  the  mean magnification  factor and  mean individual
stellar flux to get a first-order estimate:
\begin{equation}
	\hat{N}=\frac{L_{obs}}{\overline{\mu}\langle L_i\rangle} \,. \label{Nest}
\end{equation}
Thus,
\begin{equation}
\label{epsilondeltah}
	\hat{\varepsilon}_{\delta L}=\sqrt{\frac{\langle L_i^2\rangle}{\langle L_i\rangle L_{obs}}}\sqrt{\varepsilon^2_{\mu}\overline{\mu}} \, ,
\end{equation}
with   the  first   factor   in  this   formula  being   approximately
$6.02/\sqrt{L_{obs}/L_{\odot}}$   for  the   luminosity   function  of
\citet{jahreisswielen} and the variability extent
\begin{equation}
\label{epsilonmusq}
\varepsilon_{\mu}^2\equiv\frac{\sigma_{\mu}^2}{\overline{\mu}^2} = \frac{\overline{\mu^2}}{\overline{\mu}^2} - 1 \, .
\end{equation}
This quantity is therefore of our prime interest.

\subsection[]{Evaluation of $\varepsilon_{\mu}^2$}

In   calculating the variability extent $\varepsilon_{\mu}^2$  we   employ  the   method  of
\citet{neindorf},  who  improved  and  generalized  previous  works  of
\citet{deguchiwatson},  \citet{schneider1}  and \citet{schneider2}  to
make  possible the calculation  of microlensing  correlation  functions in
the case of non-zero shear. We, however, slightly modify his equations and
evaluation method for our specific needs.

Let ${\bf z}$ and ${\bzeta}$ be  the light ray positions in the lens $\cal L$ and
source $\cal S$ planes,  respectively. The normalized lens equation is
then (Kayser, Refsdal, Stabell 1986; Paczy\'nski 1986)
\begin{equation}
\label{lenseq}
	{\bzeta}=\hat{J_o}{\bf z} - \rmn{sign}(1-\kappa_c)\sum\limits_i m_i\frac{{\bf z} - {\bf z_i}}{|{\bf z} - {\bf z_i}|^2}
\end{equation}
where
\begin{equation}
	\hat{J_o}\equiv\left(\matrix{1+\gamma & 0\cr 0 & 1-\gamma\cr}\right) \label{Jodef}
\end{equation}
Here $\gamma=\gamma^{\prime}/(1-\kappa_c)$, while $\kappa_c$ is the smooth matter covergence and $\gamma^{\prime}$  is the shear, both expressed in critical units
\begin{equation}
\label{Sigmao} 
\Sigma_o\equiv\frac{c^2}{4\pi G D}
\end{equation}
and 
\begin{equation}
\label{Gammao}
\Gamma_o\equiv\frac{c^2}{4 G D} \, ,
\end{equation}
where $D$ is a reduced (angular diameter) distance 
\begin{equation}
\label{Ddef}
D\equiv\frac{D_{LS}D_{OL}}{D_{OS}} \, .
\end{equation} 
The masses  of the  microlenses $m_i$  are  given in  units  of  $M_o$ --  the
quantity which  also defines Einstein radii $z_o$ and $\zeta_o$ -- physical length units in which $\bf z$ and $\bzeta$ of~(\ref{lenseq}) are expressed -- in lens and source planes:
\begin{equation}
	z_o\equiv\sqrt{\frac{4GM_o}{c^2}\frac{1}{|1-\kappa_c|}\frac{D_{LS}D_{OL}}{D_{OS}}}
\end{equation}
\begin{equation}
\label{zeta0}
	\zeta_o\equiv\sqrt{\frac{4GM_o}{c^2}|1-\kappa_c|\frac{D_{LS}D_{OS}}{D_{OL}}} \, .
\end{equation}
In  the  case  we   consider  the  microlensing  shear  ${\bf  S}({\bf
z})\equiv\rmn{sign}(1-\kappa_c)\sum m_i({\bf z}  - {\bf z_i})/|{\bf z}
- {\bf  z_i}|^2$  --  the  second  term  in~(\ref{lenseq})  --  is  an
isotropic  random variable.  Changing the sign of  $\gamma$  has the effect of only
redefining coordinate axes and since this  is not of interest for us we
drop  the $\rmn{sign}(1-\kappa_c)$  factor  in~(\ref{lenseq}) and  use
absolute value of $\gamma$ from now on.

The magnification factor at a  point ${\bzeta}$ in the source plane
may be written in an elegant form \citep{neindorf}:
\begin{equation}
\label{neindorfmu}
\mu({\bzeta})=\frac{1}{|1-\kappa_c|^2}\int\limits_{\cal    L}   \delta
({\bzeta}-\hat{J_o}{\bf z} + {\bf S}({\bf z})) \rmn{d}^2{\bf z} \, .
\end{equation}
The average of $\mu$ is
\begin{equation}
\label{muth}
	\mu_{th}=\frac{1}{|[1-\kappa_c]^2[(1-\kappa)^2-\gamma^2]|}
\end{equation}
where  $\kappa=\pi\langle  m\rangle  n$  is  the  scaled  microlensing
optical depth with $n$ being the surface number density of microlenses
and angled brackets now meaning averaging over microlens mass distribution
$\phi(m)$.

We  consider a  Lambert disc  -- a disc of uniform surface brightness -- with
radius $R$ in $\zeta_o$ units  and total flux $I_o$. The average
value of {\it observed}  flux $\langle I\rangle$ does not depend
on the microlens  mass distribution and is only  a function of lensing
macro parameters:
\[
	\langle I\rangle=I_o\mu_{th}=\frac{I_o}{|(1-\kappa_c)^2\det{\hat{J}}|}
\]
with
\begin{equation}
\label{Jmatrix}
	\hat{J}\equiv\hat{J_o}-\kappa\hat{1}=\left(\matrix{1-\kappa+\gamma
	& 0\cr 0 & 1-\kappa-\gamma\cr}\right) \, .
\end{equation}
As  shown   in  the  Appendix A,   the  following  relation   holds  for
$\varepsilon_{\mu}^2$:
\begin{eqnarray}
\label{integraly}
\lefteqn{\varepsilon_{\mu}^2+1=\frac{\langle        I^2\rangle}{\langle
 I\rangle                               ^2}=\frac{2|\det{\hat{J}}|}{\pi
 R^2}\int\int\int\rmn{d}s^{\prime}\rmn{d}r\rmn{d}\chi\frac{1}{s^{\prime}}\times}
 \hspace{0.6cm}\\   &   &  J_1^2\left(R\frac{r   s^{\prime}}{2}\right)
 J_0\left(\gamma\frac{r^2s^{\prime}}{2}\right)                   e^{-[n
 a(s^{\prime},\chi) +i s^{\prime}\cos{\chi}]r^2/2} \, . \nonumber
\end{eqnarray}
The integration  is done from zero  to infinity in $\rho$  and $s^{\prime}$ and
from zero  to $2\pi$ in  angular variable $\chi$. Function $a(s,\chi)$  in this
formula is given by \citet{neindorf} and is described in the Appendix A.

Introducing the function
\begin{eqnarray}
\label{Bdef}
	\lefteqn{B(s, \chi)\equiv\frac{n a(s,\chi)+i s\cos{\chi}}{s}=} \hspace{0.7cm}\\
& & \frac{\kappa}{\pi\langle m\rangle}\frac{a(s,\chi)}{s}+i\cos{\chi} \nonumber
\end{eqnarray}
and performing the variable change $r\rightarrow\rho=r^2s/2$, $s^{\prime}\rightarrow s=R^2s^{\prime}/2$
we get the following integral to be evaluated:
\begin{eqnarray}
\label{thebigthing}
\lefteqn{\varepsilon_{\mu}^2+1=\frac{|\det{\hat{J}}|}{\pi}\int\limits_0^{\infty}\rmn{d}s s^{-2}\times} \hspace{0.6cm}\\
& & \int\limits_0^{\infty}\rmn{d}\rho J_1^2\left(\sqrt{\rho s}\right)J_0\left(\gamma\rho\right)\int\limits_0^{2\pi}\rmn{d}\chi e^{-\rho B\left(\frac{2s}{R^2},\chi\right)} \nonumber
\end{eqnarray}
This evaluation is done in Appendix B under the assumptions that $R\leq 10^{-2} - 10^{-3}\ll 1$ and $R\ll\gamma$.

The former is plausible, since in a cosmological situation with $D\sim 10^{28}$~cm, the length scale~(\ref{zeta0}) is
\[
\zeta_o\sim
10^{17}\left(\frac{M_o}{M_{\odot}}\right)^{1/2}\left|1-\kappa_c\right|^{1/2}\rmn{cm} \, ,
\]
while  typical   physical  sizes  of  sources  are   $\sim  10^{10}  -
10^{12}$~cm.  However, it immediately places a constraint on the microlens masses and the smooth component convergence:
\begin{equation}
\label{constraint}
M_o|1-\kappa_c|\ge 10^{-4} M_{\odot} \, .
\end{equation}
Therefore  the  results derived below  are  not  directly  applicable  to
situations where~(\ref{constraint}) is not fulfilled, which may be of
interest  when  Jupiter-mass  lenses  are  involved  or  for  detailed
investigations  of  microlensing   in  the  region  $|\kappa_c  -1|\ll
1$. Microlensing of large sources was considered numerically by
\citet{refsdalstabel1} in the zero shear  case and later  including the effect of a shear
term \citep{refsdalstabel2}.  Quite naturally,  increasing the
size  of  the  source  suppresses  microlensing-induced  fluctuations,
averaging  them  over less  correlated  regions  of the  magnification
map. However, for  sources as large as $R\sim 30$  they find values of
$\varepsilon_{\mu}\sim 0.1$ in a  range of $0\le\kappa\le 2$ and $0\le
\gamma\le 0.4$ (it was found that $\varepsilon_{\mu}^2\approx 2\kappa/R^2$ in zero shear case). This is only an order of magnitude less than the numbers we obtain below and shows, that even objects billions times less massive than the Sun could introduce noticeable variability provided they contribute to the overall compact object density (however, the time scale of this kind of variability will be much shorter).

The second assumption -- which relates $\gamma$ and $R$ -- is of a rather technical nature and does  not restrict  our scope whenever  marginal cases of zero shear are not
considered.

As shown in the Appendix B, under these assumptions the integral~(\ref{thebigthing}) may be then rewritten in the following form:
\begin{equation}
\varepsilon_{\mu}^2(\kappa, \gamma, R)=\frac{|\det\tilde{J}|}{\pi}\left[I(\kappa, \gamma)-\frac{g_3(\kappa,\gamma)}{2}\ln{R}\right] - 1 \, ,
\label{itutti}
\end{equation}
where $I(\kappa,\gamma)$ and $g_3(\kappa,\gamma)$, defined by~(\ref{Itut}) and~(\ref{g3def}), are computed numerically. The actual values of $g_3(\kappa, \gamma)$  do not  exceed $\sim 0.1  - 0.2$.  Therefore we neglect  the weak  dependence   of  $\varepsilon_{\mu}^2$  on  $R$  putting $R_o=10^{-6}$. 

\begin{figure}
\hspace{0cm}\includegraphics{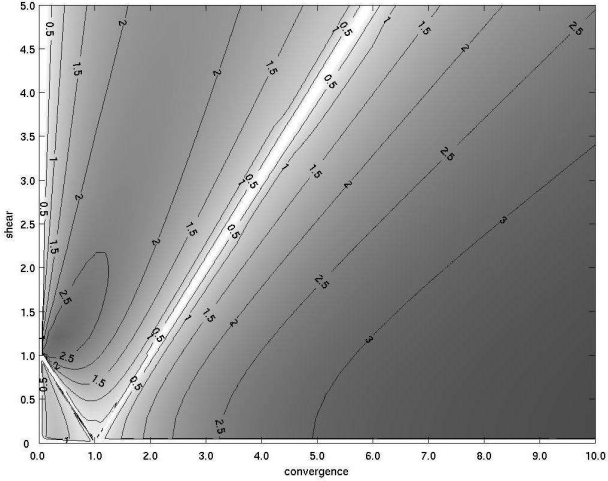}
  \caption{The contour lines of $\varepsilon_{\mu}$ as a function of convergence in microlenses $\kappa$ and shear $\gamma$.}
\label{epsilonmufig}
\end{figure}

The  contour  lines  of  constant  $\varepsilon_{\mu}$  are  shown  in
Figure~\ref{epsilonmufig}  for the  range of  parameters  $\kappa$ and
$\gamma$ present in the observed clusters. Computations near the lines
$\gamma=|1-\kappa|$ are unreliable and are therefore ignored on the graph.

\subsection{Dependence on source redshift and smooth matter contribution}

The actual values of $\kappa$ and $\gamma$ depend on both the physical
surface density at the point in  the lensing plane where the image
is formed  and the scaling parameters $\Sigma_o$  and $\Gamma_o$ given
by~(\ref{Sigmao},\ref{Gammao})  which  in turn  are  functions of  the
distance parameter  $D$~(\ref{Ddef}) and thus --  of lens and  source redshifts and
underlying cosmology. If we consider the redshift of the lens $z_l$ as
fixed, the values of convergence  and shear would depend on the source
redshift      $z_s$.     As      $\Sigma_o\propto      D^{-1}$     and
$\kappa\propto\Sigma_o^{-1}$  and  the  same  applies  to  the  shear,
$\kappa$  and $\gamma$  are directly  proportional to  $D$ and  can be
written in a simple form:
\begin{equation}
\label{kgprop}
\left(\matrix{\kappa  \cr \gamma \cr}\right)  = \left(\matrix{\kappa_o
\cr \gamma_o \cr}\right)\frac{D(z_s)}{D(z_s^o)} \, .
\end{equation}
with  $\kappa_o$  and  $\gamma_o$  being  the  convergence  and  shear
corresponding to a given redshift $z_s^o$.

For the currently favoured flat cosmological models ($\Lambda=1-\Omega_o$)
\[
D(z_s)\propto\frac{f(z_s)-f(z_l)}{f(z_s)} = 1-\frac{f(z_l)}{f(z_s)} \, ,
\]
where
\begin{equation}
\label{fdef}
f(z)=\int\limits_0^z\frac{\rmn{d}\zeta}{\sqrt{\Omega_o(\zeta+1)^3+1-\Omega_o}} \, .
\end{equation}
As $f(z)$ is  a monotonic increasing function of  its argument and
$z_s$   is   clearly  greater   than   $z_l$,   the  second   fraction
in~(\ref{kgprop})
\begin{equation}
\label{hdef}
h(z_s, z_s^o)\equiv\frac{D(z_s)}{D(z_s^o)}=\frac{1-f(z_l)/f(z_s)}{1-f(z_l)/f(z_s^o)}
\end{equation}
increases from zero at $z_s=z_l$  through unity at $z_s=z_s^o$ to some
limiting  value  $h_{\infty}$  --  determined by  $z_l$,  $z_s^o$  and
$\Omega_o$   --    when   $z_s\rightarrow\infty$. This   is   somewhat
different from considering convergence and shear as functions of $z_l$, 
in which  case there exists  an optimal lens redshift  which maximizes
the lensing  parameters. In  the case  of  varying $z_s$  the further  the
source is  the greater $\kappa$ and  $\gamma$ are. For  instance, in the
case  of Abell  370 with  $z_l=0.37$, $h_{\infty}\approx
1.7$ for  $\Omega_o=1$ and  about 1.5 when  $\Omega_o$ is only  0.2 ($z_s^o=1$) --
that  is, $\kappa$  and $\gamma$  for far  away sources  are  not much
larger than for sources at redshifts of about unity.

The behaviour  of $\varepsilon_{\mu}$ with redshift  is evidently more
complex  -  as   $\kappa$  and  $\gamma$  slide  along   the  line  of
proportionality~(\ref{kgprop})   in   $\kappa$--$\gamma$   plane   the
variance first increases  from zero at its bottom  left corner but can
then,  depending  on  $\kappa_o$  and  $\gamma_o$  cross  one  or  two
`zero-signal'  lines $\gamma=|1-\kappa|$.  Actual  $z_s=1$ convergence
and  shear in  Abell 370  and Abell  2218 for  which we  have detailed
density maps  \citep{kneib, kneib2218} cover  approximately the range  present on
figure~\ref{epsilonmufig}  therefore   there  is  no   much  point  in
discussing how $\varepsilon_{\mu}$  changes with redshift any further,
specially since  the measurements of the redshift  have been performed
for many  of the potential targets for  surface brightness variability
observations.

However,  it  is  worth  noting  the  general  pattern  of  brightness
variability  behaviour over  the area  of some  of the  most prominent
candidates for this sort  of observations - gravitational lensed
arcs.  These  objects often  consist  of two  or  more
sections  with critical  lines between  these sections and, in the case where there is no smooth matter,the variability will  be most easily  observed in the pixels  further away from the critical lines on which $\det\tilde{J}$ -- and variability -- vanish. 


When compact objects make up only a limited fraction of the lensing matter -- which is expected to be the case -- the situation is more interesting. Let  $x$ be the  compact objects  share in  the total  convergence, so that
$\kappa^{\prime}=x\kappa_{tot}$ and $\kappa_c=(1-x)\kappa_{tot}$. Then
the effective          convergence          and         shear          are
$\kappa=x\kappa_{tot}/|1-(1-x)\kappa_{tot}|$                        and
$\gamma=\gamma^{\prime}/|1-(1-x)\kappa_{tot}|$.       The       factor
$\det\hat{J}$ in $\epsilon^2_{\mu}$ is then:
\begin{eqnarray}
\lefteqn{\det\hat{J}=(1-\kappa)^2-\gamma^2=\frac{[|1-(1-x)\kappa_{tot}|-x\kappa_{tot}]^2-\gamma^{\prime
2}}{|1-(1-x)\kappa_{tot}|^2}   }   \hspace{0.1cm}\nonumber   \\  &   &
=\frac{1}{[1-(1-x)\kappa_{tot}]^2}\left\{\matrix{(1-\kappa_{tot})^2-\gamma^{\prime
2}\cr [1-(1-2x)\kappa_{tot}]^2-\gamma^{\prime 2} \cr}\right.
\label{dissolve}
\end{eqnarray}
the latter  alternative determined  by whether $\kappa_{tot}$  is less
(top) or greater (bottom) than $(1-x)^{-1}\ge 1$.

Therefore, the outer lines  of the  zero variability  signal  (those which
correspond to  $\gamma=1-\kappa$ and therefore  $\kappa < 1$)  are not
affected by the addition of  smooth matter. Other zero line positions
depend on the value of  $x$ and this dependence represents a potential
means to determine this value. We will see that only highly magnified pixels show variability detectable with present-day observational techniques -- i.e., those lying near the critical curves, and therefore for the effect to be detectable these curves should not coincide with the lines of zero variability. Thus, the condition $\kappa_{tot}\ge (1-x)^{-1}$ or, since $\kappa_{tot}$ can be determined from macrolensing modeling:
\begin{equation}
x\le 1 - 1/\kappa_{tot}
\label{condix}
\end{equation}
is in practice necessary to observe the effect. For axially symmetric clusters, the arcs which form on the second (inner) critical curves tend to have radial morphology, i.e. their dimensions along the critical curve -- and thus the number of highly magnified pixels -- are small. 

\begin{figure*}
\hspace{0cm}\includegraphics{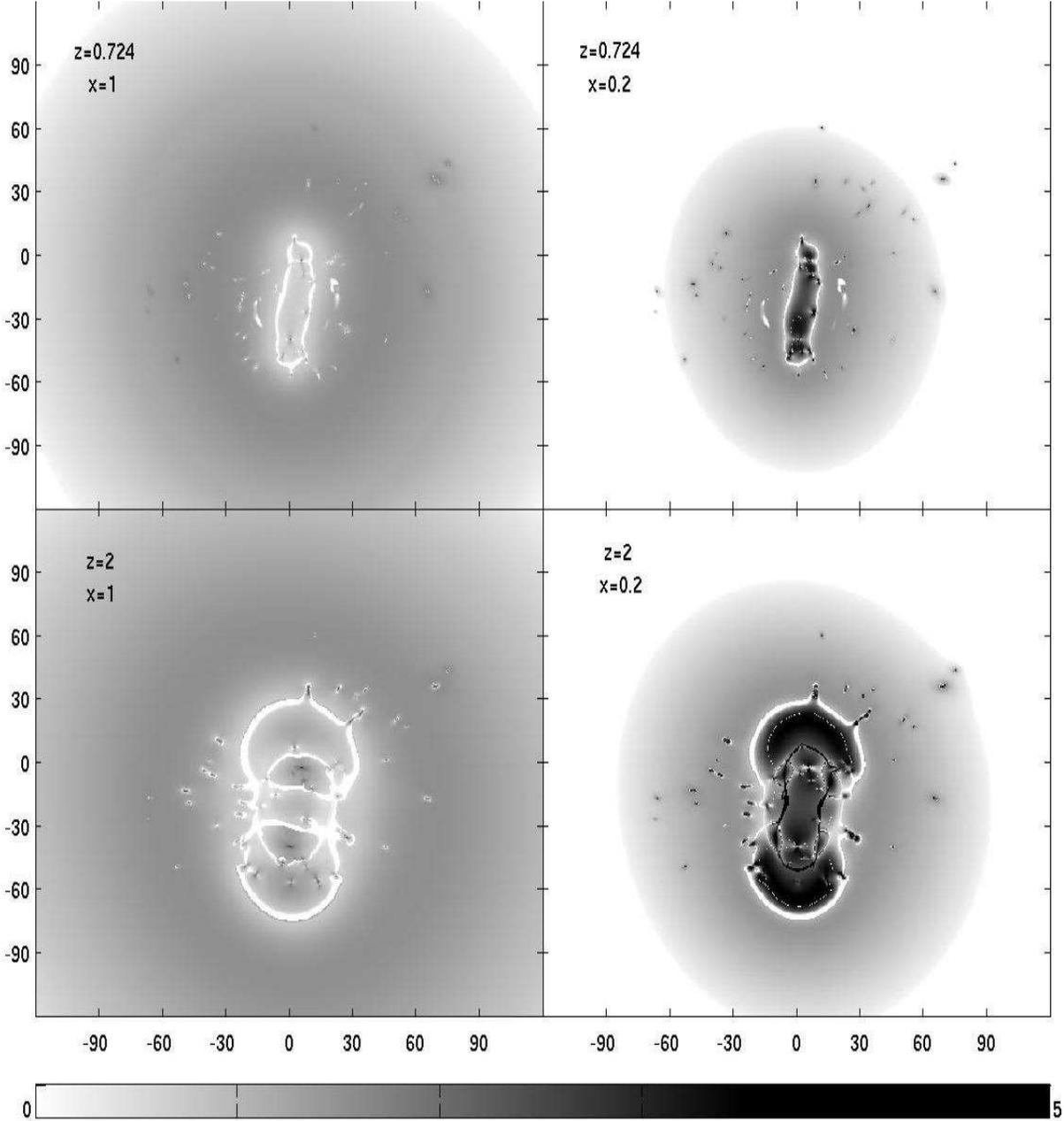}
  \caption{The  map   of  microlensing-induced  variability  parameter
  $\varepsilon_{\mu}$ over the area of the cluster Abell 370. Coordinates
  are given in arc seconds, orientation of the images as the same as in~\citet{kneib} - the north is to the top, the east is to the left.
  The  maps are given for two redshift values $z=0.724$ and $z=2$ and for two values
  of compact object  mass fraction $x=1$ and $x=0.2$ (for $x=1$ white lines of $\varepsilon_{\mu}$ coincide with the macrolensing critical lines, according to~(\ref{dissolve})).  The thick black
  lines  at $x=0.2$  correspond  to regions  with $|\kappa_c-1|\ll  1$
  where the analysis given in this paper is not applicable.}
\label{e370}
\end{figure*}

\begin{figure*}
\hspace{0cm}\includegraphics{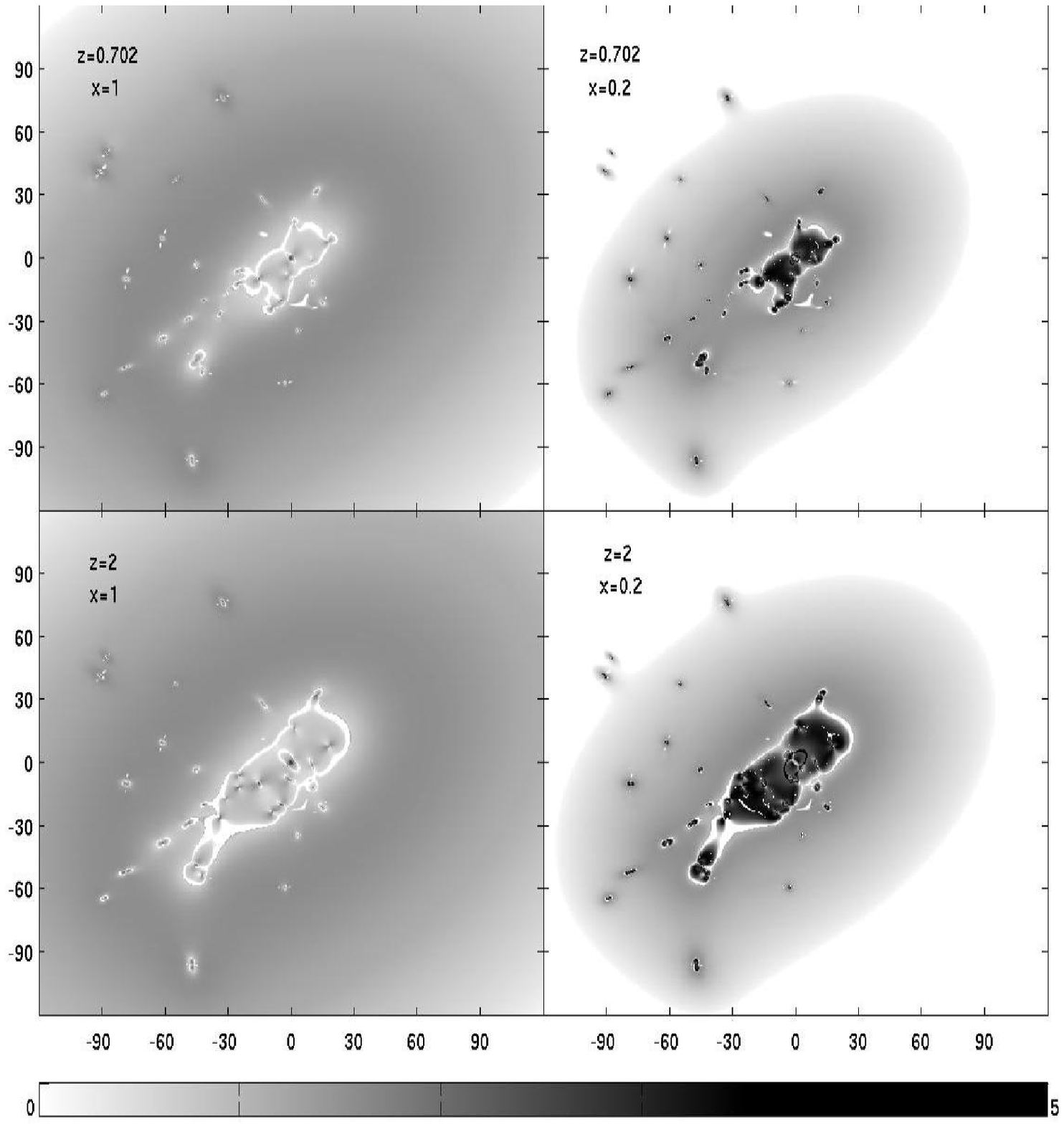}
   \caption{The  map  of  microlensing-induced  variability  parameter
   $\varepsilon_{\mu}$   over    the   area   of    the   cluster   Abell
   2218. Orientation is the same as in figure~\ref{e370}. The maps are
   given  for two  redshift values  $z=0.702$  and $z=2$  and for  two
   values of compact object mass fraction $x=1$ and $x=0.2$(for $x=1$ white lines of $\varepsilon_{\mu}$ coincide with the macrolensing critical lines, according to~(\ref{dissolve})). The thick black lines at $x=0.2$  correspond to regions with $|\kappa_c-1|\ll
   1$ where the analysis given in this paper is not applicable.}
\label{e2218}
\end{figure*}

As an example we have computed  the maps of the signal $\epsilon_{\mu}$ for
two   well-studied  clusters  -   Abell  370   and  Abell   2218  (see
\citealt{kneib2218, kneib, metcalfe} and references therein) and present
them in  figures \ref{e370} and  \ref{e2218}. These are given  for two
values of the  source redshift $z_s=2$  for both clusters  and $z_s=0.724$
and $z_s=0.702$ for Abell 370 and Abell 2218, respectively. The latter
two correspond to giant gravitational-lensed arcs seen in the clusters
while the former are given  for comparison. As new instruments -- like {\it James Webb Space Telescope} (JWST, formerly known as NGST)  -- come  into operation  they are  expected to  observe  many more
lensed galaxies  behind these clusters  and $z_s=2$ maps show  how the
signal might look  for them. Each of the maps is  given for two values
of  $x$ with  100  and 20  per  cent of  convergence contained  in
compact objects. These  values of $x$ are assumed  to be constant over
the maps.

\begin{figure*}
\hspace{0cm}\includegraphics[width=170mm, height=170mm]{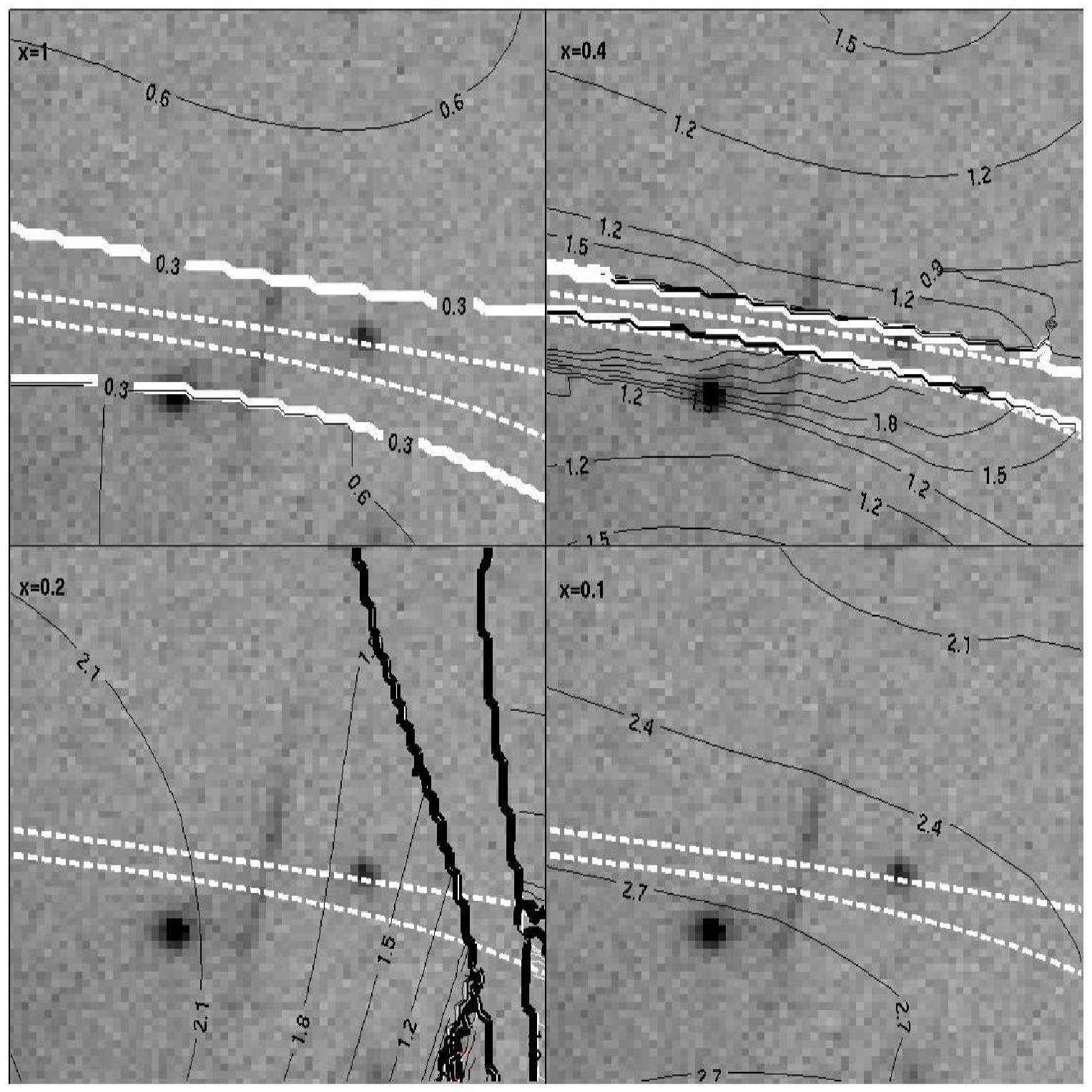}
   \caption{The contour lines of  microlensing-induced  variability  parameter
   $\varepsilon_{\mu}$ superimposed on the optical image of the radial gravitationally lensed arc ($z\approx 1.7$) seen in Abell 370 obtained with the {\it Hubble Space Telescope} \citep{kneib}. The area between the solid white lines correspond to nearly zero signal with $\varepsilon_{\mu}\le 0.3$. The dashed white lines represent the approximate location of the critical curve, while the area between the thick black lines correspond to $|\kappa_c-1|\ll 1$ where the analysis given in this paper is not applicable. Four values of compact to total mass density ratio are assumed: $x=1.0$, $x=0.4$, $x=0.2$ and $x=0.1$.}
\label{ra}
\end{figure*}

In Figure~\ref{ra} we show the contour lines of the microlensing signature $\varepsilon_{\mu}$ superimposed on the optical image of the radial arc R in Abell 370 obtained with the {\it Hubble Space Telescope} (HST) \citep{kneib}. The source redshift here is estimated to be $z_s\approx 1.7$ (see \citealt{kneib} and \citealt{smail}). The regions between the thick white lines correspond to `zero signal' lines where $\varepsilon_{\mu}\le 0.3$, while the regions between the thick black lines -- where present -- have $|\kappa_c-1|\ll 1$, where the analysis given in this paper is not applicable. Dashed white lines show the location of the critical curve.

The figure illustrates how the signal changes with varying fraction of compact objects in the overall mass budget. Perhaps in contrast to naive expectations, the signal generally increases when the density of compact objects is decreased because of the magnification effect by the smooth matter distribution. This can be understood on the basis of~(\ref{itutti}): the dependence of the variability extent on the source size is rather modest while slight changes in the smooth matter convergence $\kappa_c$ change the $|1-\kappa_c|^{-1}$ factor in the definition of effective shear and convergence of the compact matter \citep{KRS, paczynski} significantly when $\kappa_c$ is about unity which is a common place for macrolensed images of distant sources; thus convergence and shear on the $\kappa-\gamma$ plane of Fig.\ref{epsilonmufig} can assume high values. This somewhat surprising behaviour has also been discussed by \citet{schechterondissolve} while \citet{schechterwambsganss} give a detailed explanation of this effect. More important is the change in zero signal lines pattern that can be readily probed in observations and can provide, via~(\ref{dissolve}), an interesting constraint on the compact object contribution to the overall convergence determined by modeling of the lensing potential.

For images which do not lie on the critical lines observations can still be of interest for the determination of $x$ through studying the variability pattern in greater detail and comparing it to the predicted one. However, for the two clusters investigated in this paper the latter possibility remains mostly a theoretical one because of observational limitations.

\section{Observational aspects}

Let us now discuss the prospects for the detection of the considered effect. We will consider observations with {\it Hubble Space Telescope} as a reference point in this section although it will be clear that observations of this effect with HST in the two clusters under investigation is impractical. Observations with more advanced instruments, such as {\it James Webb Space Telescope} or the proposed 30-metre telescope (also known as CELT)\footnote{See {\it http://tmt.ucolick.org/} for details} could, however, be used to observe the microlensing-induced variability.

The number of photons $l$ detected in a unit time interval in a pixel from a source of observed luminosity $L_{obs}$ (uncorrected for lensing magnification) is determined by the luminosity distance to the source $D_L(z_s)$, the energy distribution in its spectrum $f_{\lambda}$ (such that $\int\rmn{d}\lambda\,f_{\lambda} = 1$) and the telescope efficiency $\eta_{\lambda}$ and diameter $d$:
\begin{eqnarray}
\lefteqn{l=\frac{d^2 L_{obs}}{16 D^2_L(z_s)}\int\rmn{d}\lambda f_{\lambda}((1+z_s)^{-1}\lambda)\eta_{\lambda}(\lambda)\frac{\lambda}{hc}}\label{photonsinpixel} \\
& = L_{obs}d^2\eta_{eff}\lambda_{eff}/16 D^2_L(z_s){hc} = \alpha L_{obs}
\nonumber
\end{eqnarray}

These photons will be accompanied by $b$ background photons. For HST Wide Field and Planetary Camera 2 (WFPC2) and actual luminosities observed in pixels of gravitationally lensed galaxies, the noise is dominated by Poissonian fluctuations in count numbers. At $z_s=1.7$ the coefficient in~(\ref{photonsinpixel}) is $\alpha\approx 5\cdot 10^{-6}\,L_{\odot}^{-1}\cdot\rmn{hour}^{-1}$ while the background level is about $b=1000$ photons in a pixel per hour (this value changes by about $1.5-2$ depending on the source heliocentric ecliptic longitude).

Let us now calculate the time $t$ required to detect the fractional change of $\beta\varepsilon_{\delta L}$ in a pixel with given microlensing parameters $\varepsilon_{\mu}$, $\overline{\mu}$ at a signal-to-noise level $Q$; $\beta$ here determines the fraction of pixels deviating from the mean at $\beta\varepsilon_{\delta L}$ level via the normal law $1-\Phi(\beta)$ and we will use $\beta=1$ for numerical estimates which is close to the optimal value. The signal $S$ is 
\begin{equation}
   S=\beta\varepsilon_{\delta L} l t \label{signalS}
\end{equation}
while the noise $N$ is determined by Poissonian fluctuations 
\begin{equation}
N=\sqrt{2bt+(2+\beta\varepsilon_{\delta L})lt} ,\label{noiseN}
\end{equation}
the factor of 2 in the latter expression comes from the fact that we need to compare two images from different epochs. In most cases $\varepsilon_{\delta L}$ can be neglected for noise estimation. Equation~(\ref{epsilondeltal}) shows that signal and noise behave similarily which, as can be easily seen, gives the following expression for the time required:
\begin{equation}
t=Q^2\frac{2\langle L_i\rangle}{\langle L_i^2\rangle\alpha}\frac{1+b/l}{\beta^2\varepsilon_{\mu}^2\overline{\mu}}
\label{exposuretime} \, .
\end{equation}
Thus, it is determined mostly by the telescope and geometry (through $\alpha$) and lensing characteristics of the pixel (through $\varepsilon_{\mu}$ and $\overline{\mu}$) while the dependence on photometry is very weak as soon as the background value is exceeded by the source surface brightness and increases inversly proportional to the latter if it is lower than the sky level. In fact, the surface brightness of a typical galaxy at $z=0$ is of order $21^m - 22^m$ per sq.arcsec and scales as $(1+z)^{-(4-p)}$ with $p$ depending on the spectrum; surface brightness is conserved in gravitational lensing. The sky background outside the Earth's atmosphere varies in the range of $22^m - 23^m$ per sq.arcsec and is about half a magnitude higher for the best terrestial observatories. Therefore, typical values of the numerator in the last fraction of~(\ref{exposuretime}) is of order 1 for nearby galaxies and grows rapidly as the redshift exceeds unity.

Combining values of coefficients in~(\ref{epsilondeltal}) and~(\ref{photonsinpixel}) we find that for observation of the radial arc in Abell 370 with HST WFPC2 the value of the first fraction in~(\ref{exposuretime}) is approximately $1.2\times 10^4\,\rmn{hours}$. The value of the variability power parameter $\varepsilon_{\mu}^2$ does not exceed $\sim 15-20$ (in fact, $\varepsilon_{\mu}^2$ changes very slowly with convergence after it exceeds approximately $15$) while the ratio of sky background to the observed arc surface brightness $b/l$ for most pixels is about $7-8$ \citep{kneib}. 

Therefore the effect can be most easily observed in pixels of high magnification $\overline{\mu}$. This value does not depend on the compact-to-smooth convergence ratio, and peaks at the critical curve. The variability $\varepsilon_{\mu}^2$, on the contrary, follows the compact matter distribution and wherever some smooth matter is present, can preserve high values at the regions of high magnification. As can be seen from~(\ref{dissolve}) this is the case when the local convergence value is greater than the inverse of the smooth matter share $(1-x)^{-1}$, or $x\le 1-1/\kappa_{tot}$; otherwise the variability zero lines coincide with the critical curves and it is not possible to get both appreciable variability and high magnification values. Somewhat ironically, the compact objects can only be observed when their mass contribution is sufficiently low.

Magnification values are determined firmly by the present-day advanced methods of mass distribution modeling in lensing clusters which proved to be accurate as well as highly and successfully predictive \citep{kneib93, ebbels}. The values of convergence for the radial arc R in Abell 370 span a range of approximately 1.3 to 1.4 and therefore the maximum values of the fraction of compact matter which produce detectable signal would be approximately 23 to 30 per cent -- values close to those suggested by studies of the Galaxy and its immediate  neighbourhood \citep{alcock, lasserre, sadoulet}. All other gravitationally lensed objects in the cluster either lie on the outer critical curves (e.g, the giant gravitationally lensed arc A0) or do not show sufficient magnification values.

For the map of variability present on Figure~\ref{ra} with $x=0.2$, the values of $\varepsilon_{\mu}^2$ on the arc are about four while the magnification varies from a few dozen to a few hundreds with a handful of pixels where $\overline{\mu}$ exceeds $10^4$. Therefore, according to eqn.~(\ref{exposuretime}), for most pixels detecting variability at signal-to-noise ratio $Q=5$ with HST would require considerable integration time of a few hundreds to a few thousands hours. However, for a few pixels these exposure times will have more reasonable values of order ten hours.

Certain reservations should be made reflecting the fact that these values are dependent on the model and in this respect the distribution of the magnifications (or the derived required exposure times) is a more robust measure. However, one should keep in mind that the critical curve is necessarily a set of points with infinite magnification and therefore the number of variable pixels is determined by the length of the arc along the critical curve (which is rather small for radial arcs) and the rate at which magnification falls off the critical curve -- i.e., graduality in convergence and shear values over the image in the critical curve vicinity.

To estimate the latter value, we can rewrite~(\ref{exposuretime}):
\begin{equation}
\label{wexpt}
t=t_o\times\frac{1}{\overline{\mu}}
\, ,
\end{equation}
where
\begin{equation}
\label{todef}
t_o\equiv Q^2\frac{2\langle L_i \rangle}{\langle L_i^2\rangle\alpha}\frac{1+b/l}{\beta^2\epsilon_{\mu}^2}
\, ,
\end{equation}
which is about $6\times 10^5\,\rmn{hours}$ for the radial arc R observed with HST.
Factor $1/\overline{\mu}$ vanishes at the critical curve, and to the first order approximation, t as a function of the coordinate $d$ orthogonal to the critical curve is
\begin{equation}
\label{tacross}
t=t_o|\nabla\frac{1}{\overline{\mu}}| d 
\, .
\end{equation}
Hence, the width of the strip along the critical line on which the required integration time $t$ is less than a given value $T$ is simply
\begin{equation}
\label{widthT}
2d=\frac{2T}{t_o|\nabla\frac{1}{\overline{\mu}}|}\ge\frac{T}{t_o}\frac{1}{|1-\kappa_{tot}||\nabla\kappa_{tot}| + \gamma^{\prime}|\nabla\gamma^{\prime}|}
\, .
\end{equation}
For the patch of the critical curve near the radial arc with $\epsilon_{\mu}^2\approx 4$, $1+b/l\approx 7-8$, $\kappa_{tot}-1=\gamma^{\prime}\approx 0.3$, $|\nabla\kappa_{tot}|\approx 2\times 10^{-3} \rmn{pix}^{-1}$, $|\nabla\gamma|\approx 3\times 10^{-3} \rmn{pix}^{-1}$ and $\beta=1$ one gets $2d\approx T/600^{\rmn{h}}$ for HST. Multiplied by the dimension of the arc along the critical curve (about five for the radial arc R in Abell 370), these give the required number of pixels. 

It is immediately clear from this estimate that since integration time of more than 100 hours is hardly possible, only local (on inter-pixel scale) stationary points in $1/\overline{\mu}$ can give detectable signal for HST images -- the above mentioned strip itself is too narrow. One would need more advanced telescopes to observe the effect, such as JWST or CELT, -- or explore other lensing clusters where strongly lensed objects on inner critical curves are seen.

At visible and near-infrared wavelengths, the sky background level expected to be observed with JWST is not much different from that with HST and therefore changes mostly come from differences in the optics and spectral band via the value of $\alpha$ in eqn.~(\ref{exposuretime}). Using the JMS sensitivity calculator\footnote{See {\it http://www.stsci.edu/jwst/science/jms/jms\_flux\_form.html}} we estimate that the exposure time required for the {\it Near-InfraRed Camera} of JWST to detect the signal will be about 15 to 20 times as short as those of HST. However, even this would require exposure times of several dozen hours as $2d\approx T/30^{\rmn{h}}$ in this case. In light of the recent discovery of a candidate $z=10$ lensed galaxy behind Abell 1835 \citep{z10}, it does not seem implausible that such ultra-deep exposures with JWST will be attempted. For the radial arc in Abell 370 that would result in about a dozen variable pixels. 

The proposed 30-metre telescope would make the prospects more optimistic. For a ground-based telescope, the sky background will be a factor of $1.5-2$ higher, and the atmosphere transparency should be taken into account. However, increase in collecting area over the HST will be enormous, and the net effect will be to reduce $t_o$ to about $8\times 10^3\,\rmn{hours}$. Of further advantage would be the use of diffraction-limited mode. Taking into account possible tracking uncertainties, an estimate for the angular resolution of $\sim 0.01$ arcsec should be considered conservative. This represents a five-fold decrease in the pixel size resulting in the ability to get closer to the critical curve. Thus, according to~(\ref{widthT}), the width of the strip around the critical curve where variability can be detected in $T$ hours integration time would make $2d\approx T/1^\rmn{h}$. Similarly, the arc length along the critical curve will be covered by five times more pixels compared to HST. This means dozens of variable pixels and potential to observe the pattern of variability change both along and across the critical curve in just a few hours long exposure!

\begin{figure}
\hspace{0cm}\includegraphics[angle=270, width=85mm]{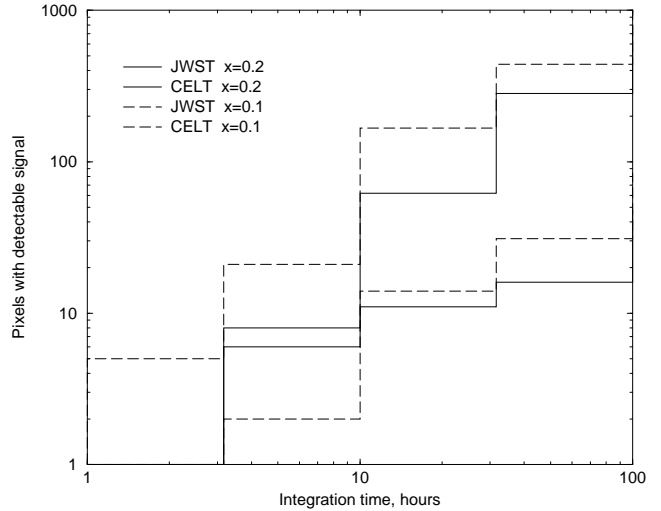}
  \caption{The number of pixels with detectable microlensing signature as a function of integration time required for the one $\varepsilon_{\mu}$ variation to be detectable at five sigma level.}
\label{histnfig}
\end{figure}

Figure~\ref{histnfig} presents the histograms of the number of pixels in the image of the radial arc seen in Abell 370 which are expected to show variability detectable at the $5\sigma$ level as a function of integration time for JWST and CELT. The actual values of the arc brightness are used to determine $t$ from~(\ref{exposuretime}) for every pixel which are then binned logarithmically in 0.5 dex wide bins. Two values of the compact matter fraction $x=0.2$ and $x=0.1$ are assumed. For JWST, the pixel size and background level are taken to be equal to those for HST WFPC2. For CELT, the background level is half a magnitude brighter while for the pixel size a value of $0.01$ arcsec is assumed.

One should bear in mind that the pixels mentioned above are variable at a detectable level, but they will only spend about one-third of the time in this `varied' state. This fraction of time can be controlled by the parameter $\beta$ but one can see -- combining~(\ref{exposuretime}), the normal distribution and the nearly linear histogram shape in the region of interest -- that the value $\beta=1$ is close to optimal.

The time scale of variations depends on the microlens masses and motion and is of order months to years for Solar masses moving with a velocity of a few hundred kilometres per second \citep{gflibataw}. Observation epochs should therefore be separated by a similar time interval.

\section{Discussion}
We have shown in the previous section that with a few exposures on JWST or CELT we expect to detect shimmering of pixels in the image of the radial gravitationally lensed arc in Abell 370 due to microlensing by compact objects in the cluster. To answer the question of how such detections should be interpreted let us now sketch a portrait of a typical event. This would also give us an insight into what sort of contaminants could mimic variability due to gravitational microlensing.

The surface brightness of the arc is about seven to eight times lower than the sky background level outside the Earth atmosphere. In a 30 hours-long exposure on JWST about 60 thousand `signal' photons will be detected in a typical pixel and around 500 thousand background photons will accompany them. Hence the noise, according to~(\ref{noiseN}), is approximately (neglecting $\beta\varepsilon_{\delta L}$) 1060 photons and five times that is 5300 photons or about 9 percent of the original flux. As $\varepsilon_{\mu}$ in this region is about 2 (see Fig.\ref{ra}C), the pixel, according to~(\ref{epsilondeltal}) contains light of approximately $4.7\times 10^4$ stars with an intrinsic luminosity of some $1.4\times 10^4\,L_{\odot}$. Given a typical observed luminosity of $2\times 10^8\,L_{\odot}$ the magnification needed is $\sim 10^4$, in accordance with the estimate of the previous section.

Nine per cent variability corresponds to $0.09\,\times\,2\times 10^8/\overline{\mu}\, L_{\odot} = 1.8\times 10^7/\overline{\mu}\, L_{\odot}$ and the average value of magnification in the map is about a hundred. In the case of the average pixel, only supernova and brightest peak nova eruptions can give the true increase of $\sim 10^5 L_{\odot}$. The most significant contaminant to the average pixel is nova eruptions. We can calculate the expected number of nova explosions in a way similar to that of \citet{ebaltz} -- for a galaxy similar to the Milky Way the rate of nova explosions is expected to be $10^{-9} - 10^{-10}$ eruptions per star per year. Eruption durations -- by which we mean the period of time novae stay above the level of interest $\sim 10^5$ -- is only a fraction $\sim 0.01 - 0.1$ of a year, long compared to the integration time while short compared to the interval between exposures. Therefore we expect about $10^{-10}$ erupted novae per star at any given exposure. The radial arc spans around three hundred pixels and therefore the expected number of stars in it is about $10^{10}$ - i.e., this is a galaxy of rather modest size. Thus, even not taking into account the multiple nature of the arc, novae are not a problem for our study. Clearly, supernova explositions in the source galaxy are even less of a problem. An additional source of contamination is supernovae in background galaxies but, with a rate of $10^{-3} - 10^{-2}$ SNe per galaxy between two exposures (see Sarajedini, Gilliland and Phillips 2000), they are not important.

However, with pixels magnified by a factor of $\sim 10^4$ which is needed to observe the microlensing variability, physical luminosity changes go down to about two thousand Solar luminosities or less, and this is about the amplitude of the brightest Mira variables in red bands. Contamination due to Miras (as well as other variable stars) in microlensing studies is usually removed by considering observations in different spectral bands. The three key signatures of microlensing origin of the variability are achromaticity, uniqueness and a symmetric form of the variations \citep{paczynski}. None of them is valid in the case considered in this paper. Achromaticity does not work for pixel lensing although in the case of low optical depth some constraints can still be applied \citep{gould96}. When the optical depth is high every star in the pixel at any given moment is subject to strong microlensing and this fact does not allow to use achromaticity constraint. For the same reason the uniqueness of the microlensing event does not work any more. With regards to the symmetry, we do not observe individual light curves in this case and therefore cannot use this constraint at all.

However, although we cannot use achromaticity for individual pairs of measurements, this property is still valid in a statistical sense. Namely, the variability extent seen in different bands is, according to~(\ref{epsilondeltal}), proportional to $\langle L_i^2\rangle / \langle L_i \rangle ^2$ -- the value which does not change much from one band to another, is closely related to the magnitude of the surface brightness fluctuations and can be determined in observations of nearby galaxies which are definitely not lensed. In contrast, (absolute) variability amplitude of variable stars is strongly dependent on the spectral band and, for instance, in Miras the change ranges from thousands Solar luminosities in K band to hundreds and even tens of Solar luminosities in bluer bands. Other variables are too faint to affect the fluxes of pixels containing thousands of stars.

Contamination due to variable stars is a more serious issue for observations with CELT. An analysis similar to the one given above, shows that typical intrinsic luminosities of pixels with variability detectable in a one hour long exposure amount to just around $500$ solar luminosities with typical variability extent of around 70 per cent or around $350$ solar luminosities. This is a range at which various variable stars may contribute to the observed variability. The only way to distinguish it from the variability due to gravitational microlensing by compact dark matter in the cluster seems to be to use the behaviour of variability from one pixel to another and across spectral bands to see whether it is consistent with physical variability or gravitational microlensing hypotheses.

One of the remaining problems is how to tell the difference between no compact matter and too much compact matter in the case of a null signal detection. More work needs to be done on this question and perhaps other effects should be considered to answer it. However, the effect considered provides us with a lot of information on the microlensing population. The Gaussian approximation seems to be a simple framework for characterizing pixel microlensing in galaxy clusters and although the implications of future observations on microlensing population are not straightforward, they can provide strong constraints on the otherwise unaccessible properties of this population.

\section*{Acknowledgments}
AVT is supported by IPRS and IPA from the University of Sydney and wishes to express his gratitude to Bernd Neindorf, Dmitry Klochkov and Mark Walker for useful discussions. GFL thanks OutKast for Hey Ya. JPK acknowledges support from Caltech and CNRS. Authors wish to thank Tim Bedding, Michael Scholz and Mike Ireland for useful explanations concerning Miras' variability and the referee for important observations.

\appendix

\section[]{Derivation of formula~(30)} 
The  derivation  presented  here  follows  very  closely  the  lines  of that by
\citet{neindorf}. We do  not include all the steps  of this derivation
which  can be  found  in the  original  work. \footnote{Please note, however, that the asymptotes for the involved functions found here differ slightly from those found in \citep{neindorf} due to some numerical error and a typo in the latter work.}

Starting      with      the      expression     for the magnification
factor~(\ref{neindorfmu})  one  has   the  following  result  for  the
observed    flux   of    a   source    with    surface   profile
$I^{\prime}({\bzeta}-{\bzeta}_o)=I_o/(\pi
R^2)\times\Theta(R-|{\bzeta}-{\bzeta_o}|)$ placed at ${\bzeta}_o$:
\begin{equation}
\label{neindorfI}
I({\bf \zeta_o})=\frac{1}{|1-\kappa_c|^2}\int\limits_{{\cal S}\times{\cal L}}\rmn{d}^2{\bzeta}\rmn{d}^2{\bf z} I^{\prime}({\bzeta}-{\bzeta}_o) \delta ({\bzeta}-\hat{J_o}{\bf z} + {\bf S}({\bf z}))
\end{equation}
Its  average  value  over  ${\bf  S}({\bf z})$  is  well-known  to  be
independent  of microlensing population  mass distribution  and source
profile (see, e.g. \citealt{schneiderbook}, Chapter 11)
\begin{eqnarray}
\label{Iav}
\lefteqn{\langle I\rangle = \frac{I_o}{|[1-\kappa_c]^2[(1-\kappa)^2-\gamma^2]|}}\hspace{-0.2cm}\\
& & =\frac{I_o}{|1-\kappa_c|^2|\det\hat{J}|}\nonumber
\end{eqnarray}
where $\hat{J}\equiv\hat{J_o}-\kappa\hat{1}$.

The value of $I^2({\bzeta_o})$ is calculated in a similar manner:
\begin{eqnarray}
\label{beginI2}
\lefteqn{I^2({\bzeta_o})=\frac{1}{|1-\kappa_c|^{4}}\int\limits_{{\cal S}^2}\rmn{d}^2{\bzeta}_1\rmn{d}^2{\bzeta}_2 I^{\prime}({\bzeta}_1-{\bzeta}_o)I^{\prime}({\bzeta}_2-{\bzeta}_o)\times} \hspace{0.5cm}\\
& & \int\limits_{{\cal L}^2}\rmn{d}^2{\bf z}_1\rmn{d}^2{\bf z}_2\delta({\bzeta}_1-\hat{J_o}{\bf z}_1+{\bf S}_1)\delta({\bzeta}_2-\hat{J_o}{\bf z}_2+{\bf S}_2) \nonumber
\end{eqnarray}
where  ${\bf S}_1={\bf  S}({\bf}z_1)$, ${\bf  S}_1={\bf S}({\bf}z_1)$.
Introducing the joint probability  function density of ${\bf S}_1$ and
${\bf S}_2$ $\varphi({\bf  S}_1, {\bf S}_2, {\bf z}_1,  {\bf z}_2)$ we
can calculate the average value of $I^2$:
\begin{eqnarray}
\label{startI2}
\lefteqn{\langle I^2\rangle =\frac{1}{|1-\kappa_c|^4}\int\limits_{{\cal S}^2\times{\cal L}^2\times{\cal R}^4}\rmn{d}^2{\bzeta}_1\rmn{d}^2{\bzeta}_2\rmn{d}^2{\bf z}_1\rmn{d}^2{\bf z}_2\rmn{d}^2{\bf S}_1\rmn{d}^2{\bf S}_2
} \hspace{0.3cm}\\
& & \delta({\bzeta}_1-\hat{J_o}{\bf z}_1 +{\bf S}_1)\delta({\bzeta}_2-\hat{J_o}{\bf z}_2+{\bf S}_2)\times \nonumber\\
& & \varphi({\bf S}_1, {\bf S}_2, {\bf z}_1, {\bf z}_2)I^{\prime}({\bzeta}_1)I^{\prime}({\bzeta}_2)\nonumber
\end{eqnarray}
Changing to the Fourier domain and making use of $\delta$~-~functions we get:
\begin{eqnarray}
\label{I2Fourier}
\lefteqn{\langle I^2\rangle =\frac{1}{(2\pi)^4|1-\kappa_c|^4}\int\limits_{{\cal R}^2\times{\cal L}^2}\rmn{d}^2{\btau}_1\rmn{d}^2{\btau}_2\rmn{d}^2{\bf z}_1\rmn{d}^2{\bf z}_2
} \hspace{0.3cm}\\
& & \tilde{I}({\btau}_1)\tilde{I}({\btau}_2)Q_2({\btau}_1, {\bf z}_1 ,{\btau}_2, {\bf z}_2) e^{-i({\btau}_1\hat{J_o}{\bf z}_1+{\btau}_2\hat{J_o}{\bf z}_2)} \, ,
\nonumber
\end{eqnarray}
where
\begin{eqnarray}
\label{Q2def}
\lefteqn{Q_2({\btau}_1, {\bf z}_1 ,{\btau}_2, {\bf z}_2)\equiv} \hspace{0cm} \\
& & \int\limits_{{\cal R}^4}\rmn{d}^2{\bf S}_1\rmn{d}^2{\bf S}_2\varphi({\bf S}_1, {\bf S}_2, {\bf z}_1, {\bf z}_2) e^{i({\btau}_1{\bf S}_1+{\btau}_2{\bf S}_2)} \nonumber
\end{eqnarray}
is the Fourier transform, or characteristic function, of $\varphi({\bf
S}_1, {\bf  S}_2, {\bf z}_1,  {\bf z}_2)$ and  $\tilde{I}({\btau})$ is
the one  of the  source profile, which is  the following  for  a uniformly
radiating disc of radius $R$ emitting a total flux $I_o$:
\begin{equation}
\label{profileFourier}
\tilde{I}({\btau})=\frac{2I_o}{R\tau}J_1(R\tau)
\end{equation}
We introduce central  and relative coordinates ${\btau}_1={\bf T}-{\bf
t}/2$, ${\btau}_2={\bf T}+{\bf t}/2$, ${\bf z}_1={\bf R}_c-{\bf r}/2$,
${\bf z}_2={\bf R}_c+{\bf r}/2$,  assume that lens positions are not
correlated and take the limit of  an infinite lens plane to obtain the
following expression ($\hat{J_o}^T=\hat{J_o}$):
\begin{eqnarray}
\label{I2viaQ}
\lefteqn{\langle I^2\rangle=\frac{1}{(4\pi)^2|1-\kappa_c|^4}\frac{1}{|\det\hat{J}|}\times}\hspace{0.5cm}\\
 & &\int\rmn{d}^2{\bf t}\rmn{d}^2{\bf r}\tilde{I}\left(-\frac{{\bf t}}{2}\right)\tilde{I}\left(\frac{{\bf t}}{2}\right)Q({\bf t}, {\bf r}) e^{-\frac{i}{2}(\hat{J_o}^T{\bf t}){\bf r}} \, ,\nonumber
\end{eqnarray}
where the $\delta$-functions in ${\bf R}_c$ and ${\bf T}$ have been utilized and for
\[ 
	Q({\bf t}, {\bf r})\equiv Q_2\left(-\frac{{\bf t}}{2}, -\frac{{\bf r}}{2}, \frac{{\bf t}}{2}, \frac{{\bf r}}{2}\right)
\]
the following expression was obtained \citep{neindorf}
\begin{equation}
\label{Qa}
	Q({\bf t}, {\bf r})=e^{-\frac{n}{2}r^2 a({\bf s})} \, .
\end{equation}
Here  $s=\frac{t}{r}$  and the  angle  $\chi$  between  ${\bf s}$  and
positive  $Ox$  ray equals  the  angle  between  ${\bf t}$  and  ${\bf
r}$. Function $a(s, \chi)$ is the mass average of $\alpha(ms, \chi)$
\[
	a(s, \chi)\equiv\int\rmn{d}m\phi(m)\alpha(ms, \chi), 
\]
the latter given by
\begin{equation}
\label{alphamsdef}
\alpha(s, \chi)\equiv\int\limits_{-\infty}^{\infty}\rmn{d}x\frac{1-e^{isx}}{x^2}f(x, \chi)
\end{equation}
Finally, the function $f(x, \chi)$ is defined as
\begin{eqnarray}
\label{fNeindorfdef}
\lefteqn{f(x,\chi)\equiv} \hspace{0cm}\\
& & \int\limits_{-\pi/2}^{\pi/2}\frac{\cos^2\phi\rmn{d}\phi}{\sqrt{(x-2\cos\chi)^2\cos^2\phi+(x\sin\phi+2\sin\chi\cos\phi)^2}} \nonumber
\end{eqnarray}
and can be expressed analytically in terms of complete elliptic integrals:
\begin{eqnarray}
\label{felliptic}
\lefteqn{f(x,\chi)=\frac{1}{u+1}\left\{\left[1+\frac{1-x\cos\chi}{u}\right]K(v)+\right.} \hspace{0.7cm}\\
 & & \Bigl.\frac{(1-x\cos\chi)(1+u)^2}{2u^2}\left[E(v)-K(v)\right]\Bigr\} \nonumber
\end{eqnarray}
with $u=\sqrt{x^2-2x\cos\chi+1}$ and $v=4u/(u+1)^2$.

The function    $\alpha(s,\chi)$     is    therefore    easily    computed
numerically. However, its overall behaviour is easily guessed from the
following two analytic asymptotes:
\begin{equation}
\label{alphalarges}
\alpha(s,\chi)\stackrel{\scriptscriptstyle s\rightarrow\infty}{\textstyle\longrightarrow}\pi s - i\frac{\pi}{2}\cos\chi
\end{equation}
valid to the accuracy of $\sim 1$ per cent at $s\ge 3 - 5$ and
\begin{eqnarray}
\label{alphasmalls}
\lefteqn{\alpha(s,\chi)\stackrel{\scriptscriptstyle s\rightarrow 0}{\textstyle\longrightarrow}\frac{\pi}{2}s^2\left(\frac{1}{2}+2\ln{2}-\tilde{\gamma}+\cos^2\chi-\ln s\right)}\hspace{1cm} \\
& & -i\pi s\cos\chi\nonumber \\
& & \approx\frac{\pi}{2}s^2\left(1.3+\cos^2\chi-\ln s\right) -i\pi\cos\chi \nonumber
\end{eqnarray}
where
$\tilde{\gamma}\equiv\lim\limits_{n\rightarrow\infty}(\sum\limits_{k=1}^n
1/k -\ln n)\approx 0.577216$ is Euler's constant.

Substituting       the      expression~(\ref{profileFourier})      for
$\tilde{I}({\bf\tau})$ into~(\ref{I2viaQ}), expanding
\[
-\frac{i}{2}(\hat{J_o}^T{\bf t}){\bf r}=-\frac{i}{2}rt[\cos\chi+\gamma\cos(\chi+2\alpha_r)]
\]
and using the change of variables ${\bf t}=r{\bf s}$ we can immediately integrate over the angular component $\alpha_r$ of ${\bf r}$ to get
\begin{eqnarray}
\label{nearlyresult}
\lefteqn{\langle                                             I^2\rangle
=\frac{2I_o^2}{\pi|1-\kappa_c|^4|\det\hat{J}|R^2}\times } \\ & &
\hspace{-0.6cm}
\int\limits_{0}^{\infty}\rmn{dr}\int\limits_{0}^{\infty}\rmn{d}s\int\limits_0^{2\pi}\rmn{d}\chi\frac{r}{s}
J_1^2\left(R\frac{rs}{2}\right)J_0\left(\gamma\frac{r^2s}{2}\right)e^{-\frac{r^2}{2}\left[n
a(s,\chi)+i s\cos\chi\right]} \nonumber
\end{eqnarray}
Thus, using  expression~(\ref{Iav}) for the average  value of observed
flux we obtain the integral~(\ref{integraly}).

\section[]{Evaluation of integral~(32)} 

To evaluate the integral in~(\ref{thebigthing}) we consider the asymptotes of $B(\sigma,\chi)$ which follow  directly from the asymptotes~(\ref{alphasmalls},~\ref{alphalarges})   of   the  function $\alpha(\sigma,\chi)$ introduced in the Appendix A:
\begin{eqnarray}
\label{Bsmalls}
\lefteqn{B(\sigma,\chi)\approx\frac{\kappa}{2}m_{\rmn{eff}}\sigma(1.3+\cos^2\chi-\frac{\langle m^2\ln m\rangle}{\langle m^2\rangle} - \ln\sigma)
} \hspace{0.4cm} \\
 & &+i(1-\kappa)\cos\chi, \hspace{2.8cm} \sigma\ll 1\nonumber \\
 & & \hspace{-1.1cm} B(\sigma,\chi)\approx B(\chi) = \kappa + i\cos\chi, \hspace{2cm} \sigma\gg 1 \label{Blarges}
\end{eqnarray}
with        $m_{\rmn{eff}}=\langle        m^2\rangle/\langle       m\rangle$
\citep{refsdalstabel1}.

At $\gamma\ge 10^{-2}$ it is convenient to split the integration over $s$ into the following five regions:
\begin{enumerate}
\item $0\le s\le\sigma_1R^2/2$, \hspace{2cm} $\sigma_1\ll 1$
\item $\sigma_1R^2/2\le s\le\sigma_2R^2/2$, \hspace{1cm} $\sigma_2\gg 1$
\item $\sigma_2R^2/2\le s\le\sigma_3\gamma^2$, \hspace{1.3cm} $\sigma_3\ll 1$
\item $\sigma_3\gamma^2\le s\le\sigma_4$, \hspace{2cm} $\sigma_4\gg 1$
\item $\sigma_4\le s$
\end{enumerate}
Due to  the assumptions on  $\gamma$ and $R$ made, in regions (i  -- iii)
the convergence  of the  integral  in $\rho$  is provided  by
$J_0(\gamma\rho)$  and $e^{-B\rho}$  while $J_1^2(\sqrt{s\rho})\approx
s\rho/4$ holds  well for all values  of $\rho$ where  the integrand is
any significant. Therefore
\begin{eqnarray}
\label{intrho123}
\lefteqn{\int\limits_0^{\infty}\rmn{d}\rho J_1^2(\sqrt{s\rho})J_0(\gamma\rho)e^{-\rho B}\approx
\frac{s}{4}\int\limits_0^{\infty}\rmn{d}\rho \rho J_0(\gamma\rho)e^{-\rho B} }\hspace{0.5cm}\\
& & \hspace{2.5cm} =\frac{s}{4}\frac{B}{(B^2+\gamma^2)^{3/2}} \, . \nonumber
\end{eqnarray}

In addition,  in regions (i) and  (iii) approximations~(\ref{Bsmalls})
and~(\ref{Blarges})   can   be    used,   respectively.   The   latter
approximation   is    even   better   in regions  (iv)    and   (v)   
but~(\ref{intrho123}) clearly  fails there.  Let us consider  the five
regions in turn:

\vspace{0.5cm}
{\it 1. $0\le s \le \sigma_1\frac{R^2}2$}

Returning to the variable $\sigma=2s/R^2$ we can write down this part
of the integral as follows:
\begin{equation}
\label{I1}
I_1=\frac{1}{4}\int\limits_0^{\sigma_1}\frac{\rmn{d}\sigma}{\sigma}\int\limits_0^{2\pi}\frac{B(\sigma,\chi)\rmn{d}\chi}{[B^2(\sigma,\chi)+\gamma^2]^{3/2}} \, .
\end{equation}
We can rewrite $B$ of~(\ref{Bsmalls}) in the following form:
\begin{equation}
\label{Bsmallsalternative}
B(\sigma,\chi)=|1-\kappa|(x\pm i\cos\chi) \, ,
\end{equation}
where
\begin{equation}
\label{xofg1}
x\equiv\frac{\kappa}{2}\frac{m_{\rmn{eff}}}{|1-\kappa|}(1.3+\cos^2\chi-\frac{\langle m^2\ln m\rangle}{\langle m^2\rangle}-\ln\sigma)\sigma
\end{equation}
and at sufficiently small $\sigma$ when the logarithm term dominates $x$ is nearly independent of $\chi$. Then
\begin{equation}
\label{I1viag1}
I_1=\frac{1}{4}\int\limits_0^{\sigma_1}\frac{\rmn{d}\sigma}{\sigma}\frac 1{|1-\kappa|^2}g_1\left(x(\sigma), \frac{\gamma}{|1-\kappa|}\right) \, ,
\end{equation}
where
\begin{equation}
\label{g1def}
g_1(x, \delta)\equiv 2\int\limits_0^{\pi}\frac{(x+i\cos\chi)\rmn{d}\chi}{\left[\delta^2+(x+i\cos\chi)^2\right]^{3/2}} \, .
\end{equation}
We have found numerically that the following approximation holds for $g_1(x,\delta)$ with an accuracy of $\le 1$ per cent:
\begin{equation}
\label{g1approx}
g_1(x, \delta)\simeq\frac{g_1(|1-\delta|, \delta)}{|1-\delta|}x, \hspace{1.2cm} x\le x_{max}=|1-\delta|
\end{equation}
$x_{max}$  stands for  the point  where $g_1$  reaches its  maximum at a
given $\delta$. This  maximum value is approximately $|1-\delta|^{-1}$
for  $|1-\delta|\le   10^{-2}$  and  should   be  calculated  directly
otherwise.

At   $|1-\delta|>\sigma_1\ln\sigma_1$   we   can   therefore   perform an
integration over $\sigma$ in~(\ref{I1viag1}) to get
\begin{eqnarray}
\label{I1final}
\lefteqn{I_1=\frac{1}{4}\frac{\kappa m_{\rmn{eff}}}{2|1-\kappa|^2}\frac{g_1\left(\left|1-\frac{\gamma}{|1-\kappa|}\right|, \frac{\gamma}{|1-\kappa|}\right)}{||1-\kappa|-\gamma|}\times
} \\
 & & \left(1.8-\frac{\langle m^2\ln m\rangle}{\langle m^2\rangle}+\ln\frac e{\sigma_1}\right)\sigma_1 \, . \nonumber
\end{eqnarray}

When $\sigma_1$ does  not obey the condition just  formulated, we just
replace    it     with    $\sigma_1^{\prime}<\sigma_1$    such    that
$\sigma_1^{\prime}\ln\sigma_1^{\prime}$      is      smaller      than
$|1-\gamma/|1-\kappa||$  and  move the  rest  of  the calculations  to
region (ii) where it is performed numerically.

The problem arises with $|1-\delta|\rightarrow 0$ but this corresponds
to   the   case   of   diverging   average   amplification   and   the
microlensing-induced variability  is expected to  drop logarithmically
to zero  in this case \citep{deguchiwatson}. Formally,  it does happen
in our calculations  - $\det\hat{J}$ in front of  the integral~(\ref{thebigthing}) make it
go to  zero linearly  in all regions  but (i)  where the ratio  of the
determinant  and  denominator of~(\ref{I1final})  tends  to a  non-zero
limit,  while  $g_1\sigma_1^{(\prime)}\ln\sigma_1^{(\prime)}$  provides
the  behaviour   expected.  However,  numerical   calculations  become
unreliable   in    this   case   and   so   we    do   not   calculate
$\varepsilon^2_{\mu}$ for  $(\kappa,\gamma)$ closer than approximately
0.03 to the $\gamma=|1-\kappa|$ lines in $\kappa$-$\gamma$ plane.

Another apparent problem with~(\ref{I1final}) seems to be present when
$\kappa\rightarrow 1$,  but this  turns out to  be a  slight technical
issue  with no real  computational consequences  and therefore  may be
called `a removable discontinuity'.

\vspace{0.5cm}
{\it 2. $\sigma_1R^2/2\le s \le \sigma_2R^2/2$}

This region is the easiest to compute:
\begin{equation}
\label{I2}
I_2=\frac{1}{4}\int\limits_{\sigma_1^{(\prime)}}^{\sigma_2}\frac{\rmn{d}\sigma}{\sigma}g_2(\sigma) \, .
\end{equation}
where
\begin{equation}
\label{g2def}
g_2(\sigma)\equiv\int_0^{2\pi}\frac{B(\sigma,\chi)\rmn{d}\chi}{\left[B^2(\sigma,\chi)+\gamma^2\right]^{3/2}}
\end{equation}
and  $B(\sigma,\chi)$ is computed numerically by interpolation of $\alpha(\sigma,\chi)$ which is computed in advance with good accuracy.

The computations of  $I_2$ provide no  problem from either  conceptual or
computational point of  view but it transpires that they are the most
time consuming part of the procedure.

\vspace{0.5cm} {\it 3. $\sigma_2R^2/2\le s\le\sigma_3\gamma^2$ }

At $s\ge  \sigma_2R^2/2$ the function $B(2s/R^2,\chi)=\kappa+i\cos\chi$ to
high accuracy and does not  depend on $s$. Therefore the integral over
$\chi$:
\begin{equation}
g_3(\kappa, \gamma)\equiv\int\limits_0^{2\pi}\frac{(\kappa+i\cos\chi)\rmn{d}\chi}{\left[(\kappa+i\cos\chi)^2+\gamma^2\right]^{3/2}} \label{g3def}
\end{equation}
(which   equals   $g_1$    introduced   above   for   $x=\kappa$   and
$\delta=\gamma$) turns out to be  a common factor  and the  integral with
respect to $s$ is elementary:
\begin{equation}
\label{I3}
I_3=\frac{1}{4}g_3(\kappa,\gamma)\ln\frac{2\sigma_3\gamma^2}{\sigma_2R^2}
\end{equation}

\vspace{0.5cm}
{\it 4 \& 5. $s > \sigma_3\gamma^2$}

In  regions (iv) and (v)  the approximation~(\ref{Blarges})  is still
valid, therefore
\begin{eqnarray}
\label{I45}
\lefteqn{
I_4+I_5=\int\limits_{\sigma_3\gamma^2}^{\infty}\frac{\rmn{d}s}{s^2}\int\limits_0^{\infty}\rmn{d}\rho J_1^2(\sqrt{s\rho})J_0(\gamma\rho)\times
} \hspace{0.5cm}\\
 & & \int\limits_0^{2\pi}\rmn{d}\chi e^{-\rho(\kappa+i\cos\chi)} = 2\pi\int\limits_{\sigma_3\gamma^2}^{\infty}\frac{\rmn{d}s}{s^2}g_4(s,\kappa, \gamma) \, ,
\nonumber
\end{eqnarray}
where
\begin{equation}
\label{g4def}
g_4(s, \kappa, \gamma)\equiv\int_0^{\infty}\rmn{d}\rho J_1^2(\sqrt{s\rho})J_0(\gamma\rho)J_0(\rho)e^{-\kappa\rho} \, .
\end{equation}
The  integrand in $g_4$  decreases exponentially  and therefore  it is
sufficient   to    perform  a numerical   integration    up   to   some
$\rho_{max}/\kappa$. The absolute value  of the residual can be easily
estimated
\begin{eqnarray}
\lefteqn{R=\left|\int\limits_{\frac{\rho_{\max}}{\kappa}}^{\infty}\rmn{d}\rho J_1^2(\sqrt{s\rho})J_0(\gamma\rho) J_0(\rho) e^{-\kappa\rho} \right|\le} \\
 & & \int\limits_{\frac{\rho_{max}}{\kappa}}\rmn{d}\rho e^{-\kappa\rho} = \frac{1}{\kappa}e^{-\rho_{max}} \, . \nonumber
\end{eqnarray}
The  value $\rho_{max}\sim  30$  turned out to be  suitable  for all  our
purposes.

For completeness  we can  write down the  integration in  region (iv),
which is done numerically:
\begin{equation}
\label{I4dummy}
I_4=2\pi\int\limits_{\sigma_3\gamma^2}^{\sigma_4}\frac{\rmn{d}s}{s^2}g_4(s,\kappa,\gamma) \, .
\end{equation}
The integration  in   region  (v)  is  accomplished   by  considering  the
asymptotic    behaviour    of     $g_4$    at    large    $s$.    The
integral~(\ref{g4def})  effectively splits  into two  -  with $\rho\le
\rho_o/s$ and  $\rho \ge \rho_o/s$  with $\rho_o\sim 1$.  Invoking the
asymptotics of Bessel functions one  can see that the second of these
integrals  is proportional  to $1/\sqrt{s}$  and represents  a leading
term   when  $s\rightarrow\infty$.   Using  the   asymptotic  formula
$J_1(x)\simeq  \sqrt{2/(\pi x)}\cos(\pi  x/2+\alpha)$ and  noting that the
$\cos$-term  oscillates rapidly for  $x\gg  1$ we  find  the  following
limiting value for $g_4$:
\[
g_4(s,\kappa,\gamma)\rightarrow\frac{1}{\pi\sqrt{s}}g_5(\kappa, \gamma) \, ,
\]
where
\begin{equation}
\label{g5def}
g_5(\kappa, \gamma)\equiv\int\limits_0^{\infty}\frac{\rmn{d}\rho}{\sqrt{\rho}}J_0(\rho)J_0(\gamma\rho)e^{-\kappa\rho} \, .
\end{equation}
This integration is again done only up to some $\rho_{max}/\kappa$.

Thus we obtain the last portion needed to compute $\epsilon^2_{\mu}$:
\begin{equation}
\label{I5}
I_5=2\pi\int\limits_{\sigma_4}^{\infty}\frac{\rmn{d}s}{s^2}g_4(s,\kappa,\gamma)\approx\frac{4}{3}\frac{g_5(\kappa,\gamma)}{\sigma_4^{3/2}} \, .
\end{equation} 

Putting all the pieces together we may now write the result:
\begin{equation}
\varepsilon_{\mu}^2(\kappa, \gamma, R)+1=\frac{|\det\tilde{J}|}{\pi}\left[I(\kappa, \gamma)-\frac{g_3(\kappa,\gamma)}{2}\ln{R}\right] \, ,
\label{tutti}
\end{equation}
where
\begin{equation}
\label{Itut}
I(\kappa, \gamma)\equiv I_1+I_2+I_4+I_5+\frac{g_3(\kappa,\gamma)}4\ln\frac{2\sigma_3\gamma^2}{\sigma_2} \, .
\end{equation}

We should note here that it is exactly the behaviour of $J_1(\sqrt{s\rho})$
-- namely the  possibility to approximate it  with $\sqrt{s\rho}/2$ --
in the  whole region of actual dependence of $B(2s/R^2, \chi)$   on $s$
that  makes the  result virtually  independent  of $R$ allowing it to appear in region (iii) only.

Considering the constants  $\sigma_i$ used  in the actual  calculations we
found that  the following values  provide the best  compromise between
the accuracy of the computations  and the time needed to perform them:
$\sigma_1=0.03$,  $\sigma_2=4$,  $\sigma_3=0.03$, $\sigma_4=30$.  This
corresponds  to the  case when $\phi(m)=\delta(m-1)$  which is
the case  actually considered. We  have checked that changes in
these constants  do not affect the  results.
\label{lastpage}
\end{document}